\documentstyle [12pt] {article}
\begin{document}
\begin{titlepage}
\begin{center}
\vspace{2cm}
\LARGE
The Observed Properties of High-Redshift Cluster Galaxies \\
\vspace{1cm}
\large
Guinevere Kauffmann \\
\vspace{0.5cm}
{\em Dept. of Astronomy, University of California, Berkeley, CA, 94720$^{
*}$}\\
\vspace{0.8cm}
\end{center}
\begin {abstract}
We use the semi-analytic models of galaxy formation developed by Kauffmann,
White \& Guiderdoni
(1993) to generate predictions for the observed
properties of galaxies in clusters and groups at redshifts between 0 and 0.6.
We examine
four different sets of cosmological initial conditions: a low-density cold dark
matter model with
and without cosmological constant,
a flat cold dark matter model and a mixed dark matter model.
These models were selected because they span a wide range in cluster formation
epoch.
The semi-analytic models that we employ are able to follow both the evolution
of the
dark matter component of clusters and the formation and evolution of the
stellar
populations of the cluster galaxies. We are thus able to generate model
predictions
that can be compared directly with the observational data.
In the low-density CDM models, clusters form at high redshift
and accrete very little mass at recent times.
Our models predict that essentially no evolution in the observed properties of
clusters
will have occurred by a redshift of 0.6, in direct contradiction with the data.
In contrast, in the MDM model, both galaxies and clusters form extremely late.
This model predicts evolution which appears to be too extreme to be in
agreement with the
observations. The flat CDM model, which is intermediate in structure formation
epoch,
is most successful.
This model is able to account for the evolution of the blue fraction
of rich clusters with redshift, the relationship
between blue fraction and cluster richness at different epochs,
and the changes in the distribution of the morphologies of cluster galaxies by
a redshift
of 0.4. In this model, galaxies have assembled most of their mass by a redshift
of 1,
but rich clusters seen today are still undergoing
considerable merging activity at redshifts as low as 0.2.
\end {abstract}
\vspace{3cm}
\small
$^{*}$ {\em present address:} Max-Planck-Institut fur Extraterrestrische
Physik,
D-85740 Garching, Germany
\end {titlepage}
\normalsize

\section {Introduction}
Clusters of galaxies have long been regarded as useful laboratories for
studying galaxy formation. Each cluster provides a sizeable sample of galaxies
that are all at the same redshift. In addition, the observed properties of
clusters at different
redshifts tells us how galaxy properties such as
morphology, luminosity, colour and metallicity evolve with time.
However, it must also be noted that
clusters are rare objects in the universe and it is likely that
physical processes such as mergers, gas stripping or starbursts have played a
rather
different role in the formation of cluster galaxies than in their field
counterparts.
A realistic theoretical model of galaxy formation in clusters must be able
to follow the assembly of the cluster over time, and it must also be able to
couple
the the star formation rate in each cluster galaxy with its interactions with
its neighbours and its environment.

Perhaps the most striking property of high-redshift clusters is the so-called
Butcher-Oemler effect -- the observation that these clusters contain
a much higher fraction of blue, star-forming galaxies than do clusters seen
today.
Recent high-resolution imaging of blue cluster galaxies, both from
the ground (Lavery \& Henry 1988,1992) and using the Hubble Space Telescope
(Couch et al 1994,
Dressler et al 1994), has begun to provide clues as to what may be causing this
enhanced level of star formation. So far the conclusion seems to be that
although
the number of interacting systems does show a substantial increase over that
observed
in present-day clusters, the majority of the blue galaxies appear to be
ordinary
star-forming spirals.
Future extension of the HST program to a larger sample of clusters and
spectroscopic follow-up work to determine which galaxies are true cluster
members
will improve the statistical confidence of these results. The hope is
that eventually the combination of multi-colour photometry, spectroscopy and
morphological
classification of a large sample of cluster galaxies at a variety of redshifts,
will
yield a detailed picture of the physical processes responsible for the
observed star formation.

In addition to understanding the processes that govern star formation, the
observed
evolution of clusters with redshift can put important constraints on
theories of galaxy and structure formation. Paper I in this series (Kauffmann
1994)
presented a simple explanation for why it is quite natural in hierarchical
clustering theories
to expect that high redshift clusters should appear more
``active'' than clusters of similar size today. It was shown that
the dark matter component of a high redshift cluster evolves much more
rapidly and undergoes more merging just prior to the epoch at which it is
observed.
Simple star formation models were introduced to illustrate how this increased
merging
activity in the dark matter component might fuel more star formation
in the galaxies themselves. The conclusion of Paper I was that it was
easy to find  plausible explanations for many of the qualitative
trends seen in the high redshift cluster data.

In this follow-up paper, we present more detailed modelling designed to yield
{\em quantitative} predictions for cluster galaxy evolution in a variety of
different cosmological models. To do this, we now follow not only the merging
history of the dark matter component of the cluster, as in Paper I, but also
the
detailed evolution of the gaseous and stellar components of each dark matter
halo in the merging ``tree''. Many of the details of this modelling have
already been
described in detail in a paper by Kauffmann, White \& Guiderdoni (1993,
hereafter KWG)
where the properties of groups and clusters at $z=0$ were studied. In this
paper, we present
results for systems at high redshift in a form that can be compared directly
with the data.

\section {Review of the Galaxy Formation Model}
In this section we present only a brief summary of the physical processes
governing
galaxy and star formation in our model. The reader is referred to KWG for
more detailed discussion.

\begin {enumerate}
\item {\em Gas Cooling}\\
Dark matter halos are modelled as truncated, isothermal spheres.
We assume that the gas relaxes to a distribution that exactly parallels that of
the
dark matter. The gas temperature can then be derived from the circular
velocity of the halo using the equation of hydrostatic equilibrium.
At each timestep in our calculation, we define
the cooling radius of a halo as the radius at which
the cooling time of the gas is equal to the age of the universe.
If the cooling radius lies outside the virial radius of the halo, the
instantaneous cooling rate is governed by the rate at which the mass of the
halo grows as a
result of the infall of surrounding matter.
Otherwise, the cooling rate is calculated from the time evolution of the
cooling radius.

\item {\em Star Formation and Feedback}\\
Once gas cools, it will fall towards the centre of the halo, forming a
dense core where stars will be able to form.
The star formation law that we choose is given by the simple equation
$ M_{*} = \alpha M_{cold}/t_{dyn}$, where $M_{gas}$ is the total mass of cold
gas
in the galaxy, $t_{dyn}$ is the dynamical timescale of the galaxy and $\alpha$
is an adjustable parameter.

Stars form with the standard stellar initial mass function given by Scalo
(1986).
The number of supernovae that are expected per solar mass of stars formed is
$4 \times 10^{-3} M_{\odot}$. The kinetic energy of the ejecta from each
supernova
is about $10^{51}$ erg. We assume that a fraction $\epsilon$ of this energy
goes
to reheat cold gas to the virial temperature of the halo, where $\epsilon$ is
taken
to be a free parameter.

\item {\em Merging of Galaxies}\\
The timescale over which the cores of halos merge once their surrounding halos
have collided is modelled using the dynamical friction timescale as given
by Binney \& Tremaine (1987):
\begin {equation} t_{dynf}= \frac {1.17 V_{c} r_{c}^{2}}
{\ln \Lambda G M_{sat}}, \end {equation}
where $V_{c}$ is the circular velocity of the ``primary'' halo, $r_{c}$ is the
virial radius of the halo, $M_{sat}$ is
the mass of the orbiting satellite and $\ln \Lambda$ is the usual Coulomb
logarithm.

In order to use equation (1) in our models, we need to specify a satellite
mass.
The appropriate choice of satellite mass is ambiguous.
Initially, most of the mass of the satellite resides in its own dark halo,
but as the satellite's orbit is eroded, a large fraction of the dark mass may
be
stripped away. In addition, angular momentum losses may shorten the dynamical
friction timescale.
Recent SPH/N-body experiments studying the assembly of galaxies in a
hierarchically
clustering universe (Navarro, Frenk \& White 1994) have shown that equation (1)
provides a good estimate of the merger timescale of gaseous cores, provided
that $M_{sat}$ is set equal to the {\em total} mass of the satellite, ie. the
mass of the baryonic + dark matter component.
In our models, we will simply take $M_{sat}=M_{baryonic}/\Omega_b$, where
$\Omega_b$ is the
fraction of the universe in baryons.

\item {\em Formation of Elliptical Galaxies and Spiral Bulges}\\
As in KWG, we assume that an elliptical galaxy is formed as the result of a
merger
between two galaxies, when the ratio of the masses of the galaxies
$M_{sat}/M_{central}$
exceeds some value $f_{ellip}$, which we keep as a free parameter.
A spiral galaxy consisting of both a disk and bulge component forms when gas
cools
onto an elliptical merger remnant.
KWG discussed the possibility that a merging between galaxies would result in a
burst
of star formation, but they
did not include this in their models. In this paper, we assume that when an
elliptical
galaxy forms, all cold gas present in it is instantaneously transformed into
stars.

\item {\em Evolution of the Stellar Populations in Galaxies}\\
We have used the spectrophotometric models of Bruzual \& Charlot (1993) to
translate the
predictions of our models into observed quantities such as magnitude and
colour,
which may be compared directly with the observational data.
\end {enumerate}

To summarize: \\
As in paper I, we use the algorithm of Kauffmann \& White (1993) to generate
Monte
Carlo realizations of the merging paths of dark matter halos from high
redshift until the present. At a given redshift, inside a given halo,
gas cools and condenses onto a {\em central galaxy} at the core of the
halo. Star formation and feedback processes take place as described above.
At a subsequent redshift, the halo will
have merged with a number of others, forming a new halo of larger
mass. All gas which has not already cooled is assumed to be shock
heated to the virial temperature of the new halo. This hot gas then
cools onto the central galaxy of the new halo, which is identified
with the central galaxy of its {\em largest progenitor}. The central
galaxies of the other progenitors become {\em satellite galaxies},
which are able to merge with the central galaxy on a dynamical friction
timescale. If  a merger takes place between two galaxies of roughly comparable
mass, the merger remnant is labelled as an ``elliptical'' and all cold gas
is tranformed into stars in a ``starburst''. Note that the infall of new gas
onto
satellite galaxies is not allowed, and star
formation will continue in such objects only until their existing cold gas
reservoirs are exhausted. Thus the epoch at which a galaxy is accreted by a
larger
halo delineates the transition between active star formation in the galaxy
and passive evolution of its stellar population.

In Paper I, we investigated two very schematic models for star formation in
cluster galaxies:
1) a model in which star formation only occurred in dark matter halos with
circular velocity comparable to those of galaxies. Once these
halos grew to the size of groups or clusters, all star formation was assumed to
stop.
2) a model in which star formation was induced by mergers taking place between
galaxy-
sized halos of roughly equal mass.
The model described above contains elements of both these toy models, but
now also treats star formation and galaxy merging in a physically realistic
and self-consistent way.

It should be noted that our models contain a number of free parameters that
control
the efficiency of star formation and supernova feedback and also the
transformation
of disk galaxies into ellipticals by merging. We fix these parameters by
requiring
that, {\em on average},  the central galaxy of a dark matter halo with circular
velocity
$V_c$= 220 km s$^{-1}$ have a luminosity, cold gas mass and morphology similar
to
that of our own Milky Way galaxy. We set $\alpha$ and $\epsilon$ to obtain a
B-band
luminosity of around $2 \times 10^{10} L_{\odot}$ and a total (HI + molecular)
gas mass of $8 \times 10^{9} M_{\odot}$. $f_{ellip}$ is set so that the
ratio of the disk to the bulge luminosity in the B band is $\simeq 5$, as is
appropriate for
a Hubble type Sb-Sc (see Simien \& de Vaucouleurs 1986). Note that the values
of
these parameters will depend on our choice of cosmology and must be reset every
time
we investigate a new model. In practice, $\alpha$ takes on values
in the range 0.05-0.2 and $\epsilon$ in the range 0.1-0.5. $f_{ellip}$ is
typically
around 0.2-0.3.

\section {Results}
\subsection {Evolution of the Blue Fraction in Clusters and Groups}

One traditional way of searching for evolution in the stellar populations of
cluster galaxies
is to measure changes in the rest-frame colours of the galaxies relative to
those of early-type galaxies at the same epoch. Butcher \& Oemler (1984)
analyzed photometry
of 33 rich clusters of galaxies with redshifts between 0.003 and 0.54.
In each cluster they selected galaxies brighter than an absolute V magnitude of
-20 and within a circular area containing the inner 30 percent of the total
cluster
population. The blue fraction $f_B$ was defined as the fraction of galaxies
whose
rest-frame B-V colours were at least 0.2 mag bluer than a typical early-type
galaxy
of the same magnitude. The blue fraction was found to increase strongly with
redshift, from an average value of around 0.04 at $z<0.1$, to a value of
0.2-0.3
at z=0.5. This analysis was extended to poorer systems by Allington-Smith et al
(1993),
who used luminous radio-galaxies as galaxy group tracers. For the poor groups,
there was no observed evolution of $f_B$ with redshift. Indeed, it was
found that by a redshift of 0.4, groups of all richnesses have similar
fractions of
blue galaxies, in contrast to the situation at z=0 where richer groups contain
fewer
blue galaxies.

In this section, we present model predictions for the evolution of $f_B$ in
groups and
clusters as a function of redshift. We have chosen to investigate four sets of
cosmological
initial conditions:
\begin {enumerate}
\item A low-density ($\Omega=0.2$) open-universe cold dark matter (CDM) model.
\item A low-density ($\Omega=0.2$) spatially flat cold dark matter model with
      cosmological constant $\Lambda=0.8$.
\item A flat CDM model with b=1.5. This normalization is inconsistent with the
amplitude of microwave background fluctuations determined by COBE, but produces
roughly the
correct abundance of rich clusters in the universe (White, Efstathiou \&
Frenk  1993).
\item A mixed dark matter model with $\Omega=1$ and a neutrino fraction
$\Omega_{\nu}=0.3$. This model is normalized to COBE.
\end {enumerate}
In all three models we take $H_0 = 50$ km s$^{-1}$ Mpc$^{-1}$ and a baryon
fraction
of 0.1. The parameters that control star formation, feedback and merging are
fixed as described in section 2. It should be noted that the epoch of group and
cluster
formation moves closer  and closer to the
present day as we go from model 1 to model 4 in the list above.
In the open model, the growth of perturbations has slowed considerably by the
present epoch
and clusters accrete very little mass at low redshift. In the MDM model,
perturbations
on small scales are erased below the neutrino Jeans length and as a result,
galaxies
only form in abundance at redshifts less than 1.

In our analysis, we define the richness $N(-19)$ to be the total number of
cluster or group members with absolute V magnitude less than -19. Our models
do not provide any information about the spatial distribution of galaxies
within
clusters, so we cannot distinguish a population of galaxies residing in the
cluster core
from a population that has
just recently been accreted. If the redder galaxies are preferentially
situated in the cluster core, as is indicated by
the observations (Thompson 1986, Couch \& Sharples 1987),
the values of $f_B$ we quote may need to be scaled down somewhat to be directly
comparable
to the observations, which tend to concentrate on the central regions of rich
clusters.
Following Butcher \& Oemler (1984), we have defined the blue fraction $f_B$ to
be the fraction
of galaxies with B-V colours at least 0.2 mag bluer than the colour of the
average elliptical
galaxy at the given redshift. We note that in our model, elliptical galaxies do
not
exhibit a colour-magnitude relation; elliptical galaxies of all luminosities
have
very similar stellar populations. It is likely that the observed
colour-magnitude relation
is due to a higher degree of metal enrichment in luminous ellipticals (Faber
1977),
something which we have neglected in our models.

In figure 1, we show the redshift evolution of
$f_B$ in a rich cluster for the four models described
above. Each point in the scatterplot represents one Monte-Carlo realization of
the
formation history of a cluster of $10^{15} M_{\odot}$. Results were calculated
for
clusters viewed at redshifts 0, 0.2, 0.4 and 0.6. We have introduced some
artificial
scatter in the redshift in the plots in order to display our results more
clearly.
We compare these results to the observational data  as given in figure 3 of
the paper by Butcher \& Oemler (1984).
As can be seen, there is essentially no evolution in $f_B$ for the two
low-density models.
The increase in blue fraction for the flat CDM model matches the observations
rather well,
but for the MDM model, the slope obtained is too steep.

One may well ask whether our
results are sensitive to the way we have normalized our model. Indeed, we find
that
the zero-point of the $f_B$ - redshift relation is very sensitive to our choice
of the
parameters $\alpha$ and $\epsilon$, which control how fast a galaxy uses up its
cold gas once
it has been accreted by a larger system, and hence also the timescale over
which its stellar
population reddens. We would thus obtain different zero points
by choosing lower or higher values
of $\Omega_b$ in our models and re-adjusting $\alpha$  and $\epsilon$ to obtain
the correct ``Milky Way'' normalization. Decreasing $\Omega_b$ means that less
gas is
able to cool in the halos. As a result, our normalization requires smaller
values of
$\alpha$  and $\epsilon$
and the blue fraction will then increase because star formation is less
efficient.
It should be noted, however, that the {\em slope} of the $f_B$-redshift
relation is
robust measure of cluster evolution in our models.
This is illustrated in figure 2, where we show how the $f_B$-redshift relation
changes
in three of  the  models for different values of $\Omega_b$. In the two CDM
models, we lower
$\Omega_b$ to 0.05 and obtain much higher blue fractions. In the MDM model, we
increase
$\Omega_b$ to 0.15 and obtain a smaller blue fraction. However, the slope does
not
change in each case.

We thus come to the conclusion that the
slope of the $f_B$-redshift relation is determined by how fast the dark matter
component of a rich clusters evolves at a given redshift and not by the
detailed
modelling of the gas physics and star formation.  Stated in another way, the
same
gasdynamical processes must operate in the same type of environment, regardless
of
the redshift.
The differences between {\em clusters of the same mass}  that we see at
different redshifts must reflect differences
in the evolutionary history of the cluster environment.
We cannot explain the rapid changes in cluster properties at redshifts as low
as 0.4
in the two low-density CDM models, simply because clusters at z=0 and z=0.4
have had much the same
history in this model. The MDM model predicts evolution that would appear to be
too rapid to
match the observations. Of the three models we have studied, the flat CDM model
clearly provides the best fit to the data.

In figure 3, we plot $f_B$ as a function of group or cluster richness, defined
by the number of
member galaxies brighter than $M_V = -19$. This scatterplot was produced by
running 200 Monte Carlo
realizations of the formation of dark matter halos with masses ranging from
$10^{13}$ to
$10^{15} M_{\odot}$. Figure 3 shows results for groups and clusters at z=0. The
results
are compared to the best-fit curve taken from figure 19 of Allington-Smith et
al (1993).
This curve was fit to the blue fractions
measured in radio groups, CfA groups and rich
clusters. The open and flat CDM models show
a strong decrease in $f_B$ with richness, but the MDM model displays no such
trend.
This may be understood as follows. In poor groups, galaxies brighter than
$M_V = -19$ are almost always central galaxies, or galaxies that
have been accreted very recently by the group and are thus still actively
forming stars.
In clusters, however, there are many bright galaxies that were accreted long
enough
ago so that they have run out of gas and reddened in colour. MDM clusters are
still
very blue because they were assembled at such low redshift.
As is the case with the $f_B$-redshift relation, the zero point of the
$f_B$-richness
relation may be adjusted by playing with the model parameters, but the overall
trend
in the blue fraction with richness remains fixed.
One way that we {\em could} alter the $f_B$-richness relation is to invoke
ram-pressure
stripping of the galactic interstellar medium by the hot gas in the
intracluster
medium. This process is likeley to be considerably more effective in rich
clusters than in poorer
environments and would certainly push the $f_B$-richness relation for the MDM
model
in the right direction.

It is known that the observational $f_{B}$ - richness relation is a reflection
of
the relationship between the colour and the morphology of a galaxy, and between
morphology
and environment, with richer environments containing a lower proportion of blue
star-forming spiral galaxies. In figure 4 we plot the spiral fraction versus
richness for
a flat CDM model. As can be seen, the same relationships between colour,
morphology and
environment occur in the models as in the data.

The most striking result in the Allington-Smith et al (1993) paper is the
evolution that
has occurred in the $f_{B}$ - richness relation by a redshift of 0.4. The high
redshift
$f_{B}$ - richness relation is essentially flat. This means that groups and
clusters
have evolved in a differential fashion, with clusters evolving much more
strongly than
poor groups. In Paper I, we predicted that the Butcher-Oemler effect would be
more
pronounced for rich clusters than for galaxy groups, simply because rich
clusters
are further out on the high-mass tail of the distribution of collapsed objects
in the universe
today. In figure 5, we show the detailed model predictions at z=0.4.
There has been essentially no evolution in the $f_B$-richness relation in the
low-density
models, as could be expected. The results for the flat CDM case agree well with
figure 19 of Allington-Smith at al. Both groups and clusters
have blue fractions of around 0.3 at z=0.4. The $f_B$-richness relation is also
flat for the MDM model, but the blue fractions in this model
are much too high to be in agreement with observations.

In summary, we have demonstrated that we can account for the difference in the
evolutionary behaviour of galaxy systems of different richnesses.
The flat CDM model again produces the best fit
to the observations.

\subsection {Colour-Magnitude-Morphology Diagrams}
There has been a long-standing debate among distant cluster observers as to
what is
causing the enhanced star formation rates in high redshift cluster galaxies.
Dressler \& Gunn (1982,1983) proposed that infalling gas-rich spiral galaxies
were
producing most of the observed star formation. Star formation in these galaxies
would then be truncated as the
interstellar medium was swept away. On the other hand,
Lavery and coworkers (Lavery \& Henry 1988;
Lavery, Pierce \& McClure 1992) have presented evidence for galaxy interactions
and
mergers being the primary cause, based on high resolution CCD imaging of the
the distant blue populations conducted from the ground. It is clear that
morphological
information will be the key to understanding the physical mechanisms driving
star formation
in cluster galaxies. Morphological classification using images made by the
Hubble Space Telescope
(HST) has now been published for galaxies in 3 clusters at $z \simeq 0.4$
(Couch et al 1994; Dressler et al 1994). Dressler et al present their data in
the form
of a colour-magnitude diagram of the cluster sample, with the morphological
type of the
galaxy indicated by a representative symbol. Spectroscopic confirmation of
cluster
membership has not yet been obtained for most of the galaxies, but the
indication is that
although the number of interacting systems has increased substantially from
what is typical of present-day clusters, most of the blue cluster galaxies are
star-forming spiral systems.

As described in section 2, star formation in our models is regulated both by
galaxy infall and by mergers. We do not have a detailed model for the
interaction
of an infalling galaxy with the hot gas of the intracluster medium. We simply
assume that the star formation rate in the galaxy will decline over some
timescale
as it uses up its own internal reservoir of cold gas.
Obviously processes such as ram-pressure stripping may accelerate this decline
in the
star formation rate. A merger between two galaxies of
similar mass results in the transformation of all the cold gas into stars in an
instantaneous ``burst'' and the production of an elliptical merger remnant.

In figures 6-8 we present colour-morphology-magnitude diagrams for three of the
cosmological models listed in section 3.1. The diagram for the model with
cosmological constant is omitted as it is very similar to that of the open CDM
model.
We show two representative clusters
of $10^{15} M_{\odot}$ at redshifts 0 and 0.4,
and we plot the rest-frame B-V colours of the cluster galaxies
against their rest-frame V absolute magnitudes.
Elliptical galaxies are plotted as filled circles, S0s as filled squares, and
spirals as 3-pronged pinwheels. Large open circles indicate galaxies which
have undergone a major merging event ($M_{sat}/M_{central} > 0.25$) in the
last gigayear before the cluster is observed.

At z=0, early-type galaxies form a narrow band in B-V colour in the diagram.
The average B-V colour of ellipticals is 0.85 in the
flat CDM model and 0.92 in the open model. This difference in
colour may be attributed both to the fact that structure forms earlier in
the open model, and  to the greater present-day age of an open universe
(16.64 Gyr, as opposed to 13 Gyr for the flat model).
Real elliptical galaxies have B-V colours that range from about 0.8 to 1.
As mentioned previously, we do not obtain a color-magnitude relation for the
ellipticals,
probably because we have not included any metal enrichment in the models.
The spiral galaxies in our model have a much broader distribution in colour and
are significantly
bluer than ellipticals or S0s. Their B-V colours range from about 0.5 to
0.8, depending on when they were accreted by the cluster and how much of their
cold gas reservoirs have been consumed by star formation. In the
the flat and open CDM model, spiral galaxies
constitute only about 20-30 \% of the total cluster population at z=0.
However, in the MDM model, spiral galaxies form 50-80 \% of the cluster
population.
This is not surprising, because we saw in the previous section that $f_B$ for
MDM clusters was much too high to match the observations.
At z=0, the number of interacting systems is typically
less than 2-3\% of the total cluster population in all three models.

In the flat CDM and MDM models, the number of interacting systems increases
markedly
at z=0.4. In addition, the spiral fraction is also enhanced. Most of the
spirals
are very blue, actively star-forming systems. Our diagrams show that it is
these
star-forming spiral galaxies that are primarily responsible for the
Butcher-Oemler effect.
In the flat CDM model, the spiral fraction is typically about 40-50 \% at a
redshift
of 0.4. In the MDM model, virtually all the galaxies are spirals.
In contrast, the open CDM model shows very little evolution in either the
number
of blue spiral galaxies or interacting systems. The colour-magnitude diagrams
at z=0 and z=0.4
are virtually indistinguishable, except for a small shift in the
colour of the elliptical population, which may be explained simply by the
passive
evolution of the stellar populations of these galaxies.

In figure 9, we present colour-morphology-magnitude diagrams for the flat CDM
model,
this time in the observer's frame. Following Dressler et al, we plot $g-r$
colours
versus apparent $r$ magnitude for four clusters of $10^{15} M_{\odot}$ viewed
at
redshift 0.4. The resemblance of these plots to figure 2 of Dressler et al
is striking, particularly for the cluster in the lower left-hand panel. The
cluster
in the lower right-hand panel has a very low spiral-fraction and only one
merging system. Its appearance is more typical of clusters at z=0. This
illustrates
that our models predict considerable scatter in the observed properties of high
redshift galaxies.
The source of this scatter is simply the variation in the evolutionary history
of a cluster
of fixed mass at a given redshift, as predicted by the Monte-Carlo algorithm.

\section {Discussion and Conclusions}
It is important to point out that the models we have explored in this paper
were developed
originally to explain the properties of groups and clusters at $z=0$ (see KWG
for more details).
The free parameters in the models are fixed so that the central galaxy of a
halo of circular
velocity 220 km s$^{-1}$ has a luminosity, gas mass and
morphology that match those of the Milky Way at the present day. It is
therefore quite remarkable
that the models are also successful at reproducing so many of the observed
trends in the
properties of groups and clusters at high redshift. In particular, we have
demonstrated that our models can account for the evolution of the blue fraction
of rich clusters with redshift, the relationship between blue fraction and
richness, both at
z=0 and at high redshift, and the changes in the morphologies of cluster
galaxies by a redshift
of 0.4.

The real key to why rich clusters are more ``active'' at high redshift
lies in in the evolutionary behaviour of the dark matter component of these
clusters.
Rich clusters at high redshift originate further out on the high-sigma tail of
density
fluctuations in the early universe than do rich clusters today.
As shown in Paper I, it follows that high redshift clusters had
a much more turbulent merging history just prior to the epoch when they
are observed as single, virialized systems of galaxies.
If the same gas-dynamical and star formation processes operate in the same type
of environment, regardless of redshift, the observed evolution of rich cluster
properties
should provide a direct measure of how rapidly structure
in the universe is evolving on cluster scales.

In this paper, we have presented the predictions of three different sets of
cosmological
initial conditions. These cosmologicals models were chosen in order to explore
a range of different cluster formation epochs. In the low-density CDM model,
the growth
of perturbations has slowed considerably by the present epoch. Clusters form at
redshifts greater than 1 and accrete very little mass at recent times.
Our models predict that essentially no evolution in the observed properties of
clusters
will have occurred by a redshift of 0.6, in direct contradiction with the data.
In contrast, in the MDM model, both galaxies and clusters form extremely late.
This model predicts evolution which appears to be too extreme to be in
agreement with the
observations. The flat CDM model is most successful. In this model, a typical
galaxy
has assembled the bulk of its mass by a redshift of 1, a group by a redshift of
0.8
and rich clusters do not assemble most of their mass until a redshift of 0.2.

\pagebreak
\Large
\begin {center} {\bf References} \\
\end {center}
\vspace {1.5cm}
\normalsize
\parindent -7mm
\parskip 3mm

Allington-Smith, J.R., Ellis, R.S., Zirbel, E.L. \& Oemler, A., 1993, APJ, 404,
521

Binney, J. \& Tremaine, S., 1987, Galactic Dynamics, Princeton Univ. Press,
Princetin, NJ

Bruzual, G. \& Charlot, S., 1993, APJ, 405, 538

Butcher, H.R. \& Oemler, A., 1984, APJ, 285, 426

Couch, W.J. \& Sharples, R.M., 1987, MNRAS, 229, 423

Couch, W.J., Ellis, R.S., Sharples, R.M. \& Smail, I, 1994, APJ, 430, 121

Dressler A. \& Gunn, J.E., 1982, APJ, 263, 533

Dressler, A. \& Gunn, J.E., 1983, APJ, 270, 7

Dressler, A., Oemler, A., Butcher, H.R. \& Gunn, J.E., 1994, APJ, 430, 107

Faber, S., 1977, In: The Evolution of Galaxies and Stellar Populations, eds
Tinsley, B.M. \&
Larson, R.B., Yale University Observatory, New Haven

Kauffmann, G. \& White, S.D.M., 1993, MNRAS, 261, 921

Kauffmann, G., White, S.D.M. \& Guiderdoni, B., 1993, MNRAS, 264, 201 (KWG)

Kauffmann, G., 19994, MNRAS, in press

Lavery, R.J. \& Henry, J.P., 1988, APJ, 330, 596

Lavery, R.J., Pierce, M.J., McClure, R.D., 1992, AJ, 104, 1067

Navarro, J.F., Frenk, C.S. \& White, C.S., 1994, preprint

Simien, F. \& de vaucouleurs, G., 1986, APJ, 302, 564

Thompson, L.A., 1986, APJ, 306, 639

White, S.D.M., Efstathiou, G. \& Frenk, C.S., 1993, MNRAS, 262, 1023
\pagebreak

\Large
\begin {center} {\bf Figure Captions} \\
\end {center}
\vspace {1.5cm}
\normalsize
\parindent 7mm
\parskip 8mm

{\bf Figure 1:} The evolution of blue fraction with redshift for the four
cosmological models
described in section 3.1. Each point represents one Monte Carlo realization of
the
merging history of a cluster of mass $10^{15} M_{\odot}$.
The solid line is a fit to the observational data taken from figure 3 of
Butcher \& Oemler (1984).

{\bf Figure 2:} An illustration of the shift in the $f_B$-redshift relation for
different
values of $\Omega_b$ in the models . In all three panels, filled circles are
for models with
$\Omega_b =0.1$. In the two CDM models, filled triangles are for models with
$\Omega_b=0.05$.
In the MDM model, the filled triangles are for a model with $\Omega_b$=0.15.

{\bf Figure 3:} The blue fraction versus richness relation for clusters and
groups at
$z=0$. The blue fraction is plotted against $\log N[-19]$, where $N[-19]$ is
defined
to be the number of galaxies in the group or cluster brighter than $V=-19$.
The solid curve is the fit to the observational data as given in figure 19
of Allington-Smith et al (1993).

{\bf Figure 4:} The spiral fraction versus richness relation for a flat CDM
model.

{\bf Figure 5:} The blue fraction versus richness relation for clusters and
groups at
$z=0.4$.
The solid curve is the fit to the observational data at $z=0$ as given in
figure 19
of Allington-Smith et al (1993).

{\bf Figure 6:} Rest-frame B-V versus V colour-morphology-magnitude diagrams
for
galaxies in clusters of $10^{15} M_{\odot}$ in the open CDM model.
The top two panels show two representative clusters
at $z=0.4$ and the bottom panels show two clusters at the present day.
Elliptical galaxies are plotted as filled circles, S0s as filled squares and
spirals
as 3-cornered pinwheels. Large open circles indicate galaxies that have
undergone
a major merger in the past Gyr.

{\bf Figure 7:} Rest-frame B-V versus V colour-morphology-magnitude diagrams
for
galaxies in clusters of $10^{15} M_{\odot}$ in the flat CDM model.

{\bf Figure 8:} Rest-frame B-V versus V colour-morphology-magnitude diagrams
for
galaxies in clusters of $10^{15} M_{\odot}$ in the MDM model.

{\bf Figure 9:} Observed-frame $g-r$ versus $r$ colour-morphology-magnitude
diagrams for
four different $10_{15} M_{\odot}$ clusters seen at z=0.4 in the flat CDM
model.
\end {document}